\documentclass[twocolumn,showpacs,amsmath,amssymb,superscriptaddress]{revtex4}
\usepackage[dvips]{graphicx}
\usepackage{graphics}
\usepackage{dcolumn}
\usepackage{bm}
\usepackage{bbm}
\usepackage{amsmath}
\usepackage{amssymb}

\begin{document}
\title{Reconstruction of condensed magnetoexciton  droplet  in  a trap 
in strong magnetic fields}
\draft
\author{S.-R. Eric Yang}
\affiliation{Department of Physics, Korea University, Seoul 
136-701, Korea}
\author{J. Yeo}
\affiliation{Department of Physics, Konkuk University, Seoul 143-701, Korea}
\affiliation{School of Physics and Astronomy, University of Manchester, Manchester M13 9PL U. K.}
\author{S. Han}
\affiliation{Department of Physics, Korea University, Seoul 
136-701, Korea}
\affiliation{Department of Physics,
Boston University, Boston, MA 02215, USA
}

\begin{abstract}
We investigate theoretically the Bose-Einstein condensation of trapped magnetoexcitons
in  a two-layered system with one layer
containing electrons and the other layer containing holes.
We have studied the spatial variations  of the condensate density in
the droplet of electrons and holes.
We find that the shape of the electron and hole  densities may change due to the competition between 
repulsive electron-electron/hole-hole interaction and confinement potential of the trap.
Our mean field calculations show that when the confinement strength is strong enough  the condensate  density
is peaked at the edge of the droplet, and as the confinement strength weakens
the  condensate density displays one inner peak  and another peak
at the edge.  For much weaker confinement potential 
the condensate  density may display one  broad peak.

\end{abstract}

\thispagestyle{empty}
\maketitle

\section{Introduction}

The Bose-Einstein condensation (BEC) of
excitons is more likely to occur when (a) electrons and holes are spatially 
separated, (b) a strong magnetic field is present, and (c) bosons are confined  in a
trap.  When electrons and holes are spatially separated exciton lifetime increases.
The lifetime depends on the wavefunction overlap between electrons and holes.
In coupled quantum wells
the wavefunction overlap  can be controlled by an external
electric field applied perpendicular to the quantum wells.
It was shown theoretically that
in such a system 
excitonic condensate can be realized in strong magnetic fields  nearly at all filling factors
when the separation between the electron and hole layers is less than the magnetic length \cite{YM}.  
Only for large
separations fractional quantum Hall states
or Wigner crystal states may be realized.
Different aspects of excitonic BEC in two dimensional systems in strong magnetic fields
have been investigated over the years\cite{KH, paquet,kor,mos,LY,YF,BR,YM,butov,Chen,Ber}.
As the observation of atomic BEC
demonstrates \cite{wieman} the conditions for BEC are improved if bosons are confined  in a
trap.  Several groups have recently 
investigated excitonic  BEC in a trap in the absence of a
magnetic field.  Experimentally the possibility of excitonic  BEC  
was explored in traps of double quantum wells \cite{neg,Butov,Snoke}.  
Theoretically exciton BEC in a trap was investigated in a mean 
field theory including electron-electron and hole-hole interactions\cite{zhu}.
Signatures of BEC in angular distribution of photoluminescence have been also explored theoretically\cite{Kee}.

In this paper we investigate theoretically the BEC of excitons when
both a strong magnetic field and a trap are present in double quantum wells.
The magnetic field is applied perpendicular to the layers and the confinement potential
traps particles laterally, i.\ e.\ in the layers.
We consider the {\it strong magnetic field} limit where  only the states in the lowest
Landau level are relevant.
The trap potential will split the degenerate lowest Landau level states. 
What is interesting about strong magnetic field 
limit is that the particle density cannot exceed a certain value  $1/2\pi\ell^2$,
where $\ell$ is the magnetic length.
In bulk 2D this density is achieved when the lowest Landau level if completely filled, 
i.\ e.\ when the Landau
level filling factor is one.
A strong confinement potential pushes particles in each layer to the center of the potential and 
the particle density 
takes  the largest possible value   
except near the edge of such a  droplet. 
Similar effect is not present in the absence of a strong magnetic field.
We will call this state 
the  maximum density droplet (MDD). 
When the confinement potential becomes weaker the particle density can take a smaller value at some place in 
the droplet.
Such a reconstruction for MDD 
\cite{Yang1,Yang2,Yang3}
in electron single dots was  investigated both theoretically and experimentally  
by several groups in  strong magnetic fields \cite{Yang1,Yang2,Yang3,As,Oo,Kl,So,Re,Ch,Ka}.
The purpose of this paper is to investigate how a reconstruction of the   MDD affects   the condensed  magnetoexcitons.

We have explored this problem within a  BCS  theory.
In the  MDD  condensed   magnetoexcitons are found near the edge of 
the droplet, see Fig. \ref{fig:or_1}.
We find that as the strength of the confinement potential weakens 
this MDD  becomes unstable and
density depletion starts to occur at the interior of the droplet. 
In this state condensed   magnetoexcitons are found both in the interior of the droplet and 
near the edge of the droplet, see Fig.\ref{fig:or_2}.
When the confinement potential is weakened
further
the density depletion of electrons and holes increases.  This behavior is similar to how the electron MDD
reconstructs in electron single dots \cite{Yang3}.   
However, we find that, 
unlike the electron single dot case, where the number of electrons depleted  increases 
discontinuously from one, two, and etc., \cite{Yang3},
the density depletion increases
continuously with the decrease in the strength of the confinement potential.  
When density changes from the uniform state  are more severe the order parameter
may display one  broad peak.

This paper is organized as follows. In Sec.II we describe our model and its Hamiltonian.
A gap equation for the  condensate is derived in Sec.III.
In Sec.IV we show solutions of this gap equation.
Conclusions are given in Sec.V.

\section{Model}

An artificial trap for excitons 
in double quantum wells can be set up by changing locally the well width on one side of
double quantum well \cite{zhu}. An in-plane random potential may also provide
a local potential minimum which can host electrons and holes\cite{Butov,Snoke}.
A parabolic  potential may be created by applying inhomogeneous stress\cite{neg}.
Another possibility of a trap is self-assembled  quantum dots\cite{Today}.

We take a simple model for the confinement potential of such traps and 
consider  electrons and holes to be confined in their respective layers  by parabolic quantum dot 
potentials, described by a parameter $\Omega$:
$\frac{1}{2}m_e\Omega^2r^2$ and $\frac{1}{2}m_h\Omega^2r^2$, where 
$m_{e}$ and $m_{h}$ are the electron and hole masses.
A magnetic field $B$ is applied perpendicular to the 2D layers.
(Hereafter the perpendicular direction will be taken to be the z-axis).
We confine our attention here to the strong magnetic field limit, where
$\Omega/\omega_{c,\sigma}\le 1$, with $\omega_{c,\sigma}=eB/m_{\sigma}c$
and $\sigma=e,h$.
We assume that the magnetic field is so strong that  both electron and hole systems are {\it spin-polarized}.
In this limit the symmetric gauge single-particle eigenstates are
conveniently classified by a Landau level index $n$ and an angular 
momentum index $m$.   Here we consider the strong field limit so $n=0$ and $m=0,1,2,...$
Electrons with angular momentum $m$ have a wavefunction
\begin{equation}
\label{wavef}
\phi_m ({\bf r})=\frac 1 {\sqrt{2\pi \ell^2 2^m m!}}\Big(\frac {\bar z} \ell 
\Big)^m e^{-|z|^2/4\ell^2},
\end{equation}
while holes with angular momentum $-m$ have a wavefunction
\begin{equation}
\phi_{-m} ({\bf r})=\frac 1 {\sqrt{2\pi \ell^2 2^m m!}}\Big(\frac  z \ell 
\Big)^m e^{-|z|^2/4\ell^2}.
\end{equation}
Here $z=x+iy$ and $\ell=\sqrt{\hbar c/eB}$.
Note that these wavefunctions are independent
of electron and hole masses.
An electron with angular momentum $m$ has a wavefunction 
$\phi_m(\mathbf{r})\sim e^{-im\phi}$ rotating 
counter clockwise.
A hole  with angular momentum $-m$ has a wavefunction 
$\phi_{-m}(\mathbf{r})\sim e^{im\phi}$ rotating  clockwise. 
To stress the similarity between our approach and  the treatment of superconductivity
we  will {\it rename electrons and holes as pseudo spin-up and -down particles}.
The single particle
orbitals in the $n=0$ level have energies 
$\epsilon^{\sigma}_{m}=\hbar\omega_{c,\sigma}/2+\gamma_{\sigma}(m+1)$,
where $\gamma_{\sigma}=m_{\sigma}\Omega^2\ell^2$, 
$\omega_{c,\sigma}=eB/m_{\sigma}c$,  and $\sigma=\uparrow, \downarrow$ is the pseudospin index.
The term $\gamma_{\sigma}(m+1)$
represents the {\it splitting} of the degenerate states of the lowest Landau level by the potential of the trap.
The electron and hole creation operators are
$c^\dagger_{m\uparrow}$ and $c^\dagger_{-m\downarrow}$.

We will derive a gap equation for
BEC of two-dimensional magnetoexcitons \cite{yang} in a trap described above. 
Here we are
particularly interested in the competition between the confinement 
potential and the repulsive
particle-particle interactions.   We thus include electron-electron 
and hole-hole interactions
in the Hamiltonian explicitly.
Our system then consists of  pseudo spin-up and -down particles 
in two dimensions under a strong magnetic field.
The  Hamiltonian of the system is:
\begin{equation}
{\mathcal H}={\mathcal H}_{\mathrm e}+{\mathcal H}_{\mathrm h}+{\mathcal H}_{\textrm {e-h}}
+{\mathcal H}_{\textrm {h-e}},
\end{equation}
where
\begin{eqnarray}
{\mathcal H}_{\mathrm e}&=&
\sum_{m} \varepsilon^{\uparrow}_{m}c^\dagger_{m\uparrow}
c_{m\uparrow} \\
&+&\frac 1 2 \sum_{m_i,m^\prime_i}
V_s(m^\prime_1, m^\prime_2, m_1, m_2 ) \; c^\dagger_{m^\prime_1\uparrow}
c^\dagger_{m^\prime_2\uparrow}c_{m_2 \uparrow} c_{m_1\uparrow },
\nonumber \\
{\mathcal H}_{\mathrm h}&=&
\sum_{m} \varepsilon^{\downarrow}_{ -m}c^\dagger_{-m\downarrow }
c_{-m\downarrow} \\
&+&\frac 1 2 \sum_{m_i,m^\prime_i}
V_s (-m^\prime_1,
-m^\prime_2 ,-m_1,
-m_2) \nonumber\\
&&\times c^\dagger_{-m^\prime_1\downarrow}
c^\dagger_{-m^\prime_2\downarrow}
c_{-m_2\downarrow }c_{-m_1\downarrow } \nonumber 
\end{eqnarray}
are respectively the Hamiltonian for the electrons and the holes, and
\begin{eqnarray}
{\mathcal H}_{\textrm{e-h}}&=&-\frac 1 2 \sum_{m_i,m^\prime_i}
V_d(m^\prime_1,-m^\prime_2, m_1 ,-m_2)\nonumber\\
&&\times c^\dagger_{m^\prime_1\uparrow}
c^\dagger_{-m^\prime_2\downarrow}c_{-m_2\downarrow } c_{m_1\uparrow } ,
\\
{\mathcal H}_{\textrm{h-e}}&=&-\frac 1 2 \sum_{m_i,m^\prime_i}
V_d(-m^\prime_1, m^\prime_2,-m_1, m_2)\nonumber\\
&& \times c^\dagger_{-m^\prime_1\downarrow}
c^\dagger_{m^\prime_2\uparrow}c_{m_2\uparrow } c_{-m_1\downarrow }
\end{eqnarray}
describe the electron-hole interactions. Here $\sum_{m_i,m^\prime_i}$ stands for
$\sum_{m_1,m_2,m^\prime_1,m^\prime_2}$ and 
$V_d$ is defined to be positive so we put 
minus signs in front of $1/2$ in ${\mathcal H}_{\textrm{e-h}}$ and 
${\mathcal H}_{\textrm{h-e}}$.  
The  matrix elements for intra dot electron-electron/hole-hole and 
inter-dot electron-hole interactions are, respectively,  
\begin{eqnarray}
&&V_s(m^\prime_1,m^\prime_2, m_1, m_2)=\int d^2{\bf r}_1
\int d^2{\bf r}_2\;V_s ({\bf r}_1-{\bf r}_2) \nonumber\\
&&\quad\quad\quad\times\phi^*_{m^\prime_1} ({\bf r}_1) 
\phi^*_{m^\prime_2} ({\bf r}_2)
\phi_{m_1}({\bf r}_1)\phi_{m_2}({\bf r}_2)
\end{eqnarray}
and 
\begin{eqnarray}
&&V_d(m^\prime_1,m^\prime_2, m_1, m_2)=\int d^2{\bf r}_1
\int d^2{\bf r}_2\;V_d ({\bf r}_1-{\bf r}_2) \nonumber\\
&&\quad\quad\quad\times \phi^*_{m^\prime_1} ({\bf r}_1)
\phi^*_{m^\prime_2} ({\bf r}_2)
\phi_{m_1}({\bf r}_1)\phi_{m_2}({\bf r}_2).
\end{eqnarray}
Note that $V_s({\bf r})$ is just the Coulomb interaction and
$V_d(\vec{r})=\frac{e^2}{\epsilon\sqrt{ r^2+d^2}}$, where $d$ is the {\it inter layer distance} and 
$\epsilon$
is the dielectric constant of the semiconductor.
One can show that
$V_d(m^\prime_1,-m^\prime_2, m_1 ,-m_2)=V_d(-m^\prime_2,m^\prime_1, -m_2 
,m_1)$.

\section{Gap equations and self energies}

In a bulk 2D when the filling factor  $\nu\neq 1$  
the excitonic  condensate is described well by the mean field BCS state
\begin{equation}
|\Psi\rangle=\prod_{m=0}^{\infty}(u_m+v_m c^\dagger_{m\uparrow} 
c^\dagger_{-m\downarrow})|0\rangle.
\label{BCSstate}
\end{equation}
This state represents  a {\it uniform} density 
condensate  with constant  the particle occupation number, 
$v_m^2=\nu$.
Small system exact diagonalization studies\cite{YM} indicate the overlap between
this mean field uniform BCS and exact groundstate wave functions is close to one for small values of $d/\ell$.
Also a trial wavefunction leads to identical physical results\cite{YM} as mean field field results\cite{YF}.

We apply the same trial wavefunction to describe the excitonic BEC in a trap\cite{zhu}.
The expectation values of the different terms in the Hamiltonian with
respect to $|\Psi\rangle$ may be conveniently
evaluated by using
\begin{eqnarray}
c_{m\uparrow}&=&u_{m}\alpha_{m\uparrow}+v_{m}\alpha^{\dagger}_{-m\downarrow}\nonumber\\
c^{\dagger}_{m\uparrow}&=&u_{m}\alpha^{\dagger}_{m\uparrow}+v_{m}\alpha_{-m\downarrow}\nonumber\\
c_{-m\downarrow}&=&u_{m}\alpha_{-m\downarrow}-v_{m}\alpha^{\dagger}_{m\uparrow}\nonumber\\
c^{\dagger}_{-m\downarrow}&=&u_{m}\alpha^{\dagger}_{-m\downarrow}-v_{m}\alpha_{m\uparrow}\nonumber\\.
\end{eqnarray}
The $\alpha$'s satisfy anticomutation relations.  The BCS state $|\Psi\rangle$
represents the vacuum state:
$\alpha_{m\uparrow}|\Psi\rangle=0$ and $\alpha_{m\downarrow}|\Psi\rangle=0$.
Using the usual Hartree-Fock (HF) decoupling we find the expectation values
\begin{eqnarray}
\langle \mathcal{H}_{\rm e}\rangle &=&\sum_{m=0}^{m_c}\{\epsilon_m^{\uparrow}+\frac{1}{2}\sum_{m'}[V_s(m,m',m,m')\nonumber\\
&-&V_s(m,m',m',m)]v_{m'}^2\}v_m^2
\end{eqnarray}
\begin{eqnarray}
\langle \mathcal{H}_{\rm h}\rangle&=&\sum_{m=0}^{m_c}\{\epsilon_{m}^{\downarrow}+\frac{1}{2}\sum_{m'=0}^{m_c}
[V_s(-m,-m,-m,-m')\nonumber\\
&-&V_s(-m,-m',-m',-m)]v_{m'}^2\}v_{m}^2
\end{eqnarray}
\begin{eqnarray}
\langle \mathcal{H}_{\textrm{e-h}}\rangle &+&\langle \mathcal{H}_{\textrm{h-e}}\rangle=\nonumber\\
&-&\sum_{m,m'=0}^{m_c} V_d(m,-m,m',-m')v^2_mv^2_{m'}\nonumber\\
&-&\sum_{m,m'=0}^{m_c} V_d(m,-m,m',-m')u_mv_mu_{m'}v_{m'}. \nonumber\\
\end{eqnarray}
We can derive  gap equations by minimizing the total energy
$\langle \mathcal{H}_{\rm e}\rangle+\langle \mathcal{H}_{\rm h}\rangle+
\langle \mathcal{H}_{\textrm{e-h}}\rangle +\langle \mathcal{H}_{\textrm{h-e}}\rangle$:.
\begin{eqnarray}
\Delta_m &=&\frac{1}{2}\sum_mV_d(m,-m,m^\prime,-m^\prime)
\frac{ 
\Delta_{m'}}{\sqrt{\Delta^2_{m'}+(p_{m'}/2)^2}},\nonumber\\ 
\label{gapeq1}
\end{eqnarray}
where
\begin{eqnarray}
p_m &=&\epsilon^{\uparrow}_{HF}(m)+\epsilon^{\downarrow}_{HF}(m)-2\mu^{+}. 
\label{gapeq2}
\end{eqnarray}
It is useful to know
\begin{eqnarray}
\Delta^2_{m}&=&\left(\frac{p_{m}}2\right)^2\left[\frac{1}{(1-2v^2_{m})^2}-1
\right], \\
v^2_{m}&=&\frac{1}{2}  \left(    1-   \frac{p_{m}/2}{   
\sqrt{\Delta^2_{m}+(p_{m}/2)^2}  }   \right).  
\end{eqnarray}
Here we define the chemical potential $\mu^+\equiv
(\mu_\uparrow+\mu_\downarrow)/2$. 
In solving the gap equation we must determine $\mu^{+}$ self-consistently.
The occupation number of single particle states is $v_m^2$.
The  HF effective single particle energy
of a  particle with pseudo spin $\sigma$ is 
\begin{eqnarray}
\epsilon^{\sigma}_{HF}(m)=\epsilon^{\sigma}_{m}+\Sigma_{H,s}(m)+\Sigma_X(m)+\Sigma_{H,d}(m),
\label{HFenergy}
\end{eqnarray}
where 
$\epsilon^{\sigma}_{m}$ is the single particle confinement potential energy
and
\begin{eqnarray}
\Sigma_{H,s}(m)&=&\sum_{m'}V_s(m, m^\prime, m, m^\prime)v^2_{m'},\nonumber\\ 
\Sigma_X(m)&=&-\sum_{m'}V_s(m, m^\prime, m^\prime, m)v^2_{m'},\nonumber\\
\Sigma_{H,d}(m)&=&-\sum_{m'}V_d(m,-m,m^\prime,-m^\prime)v^2_{m'}.
\end{eqnarray}
The physical meaning of each term is as follows.  Consider, for example, the  HF effective single particle energy
of a pseudo spin-up particle.
The first and second terms then represent the Hartree and exchange self energy corrections
due to the presence of other pseudo spin-up particles.  The third term represents
the Hartree self energy correction due to the presence of pseudo spin-down  particles.
Note that when the inter layer distance $d=0$ then $\Sigma_{X}(m)=\Sigma_{H,d}(m)$ since 
$V_s(m, m^\prime, m^\prime, m)=V_d(m,-m,m^\prime,-m^\prime)$.

\section{Results}

We present numerical solutions of the gap equations, Eqs.~(\ref{gapeq1})-(\ref{gapeq2}), 
for a trap with  the
electron layer and hole layer 
separated in the z-direction by a distance $d=0.5\ell$. For this inter layer distance 
BEC is expected  in bulk 2D systems for filling factors less than one \cite{YM}.  
The spread of the wave functions of electrons and holes in the
z-direction is assumed to be
negligibly small.   We investigate  the ideal case with $T=0$ and equal electron and hole masses  $m_e=m_h=0.067m_0$.
In our study we have equal number of electrons and holes, $N_e=N_h=N$.

\subsection{Test of the mean field theory}

\begin{figure}
\includegraphics[width=0.45\textwidth]{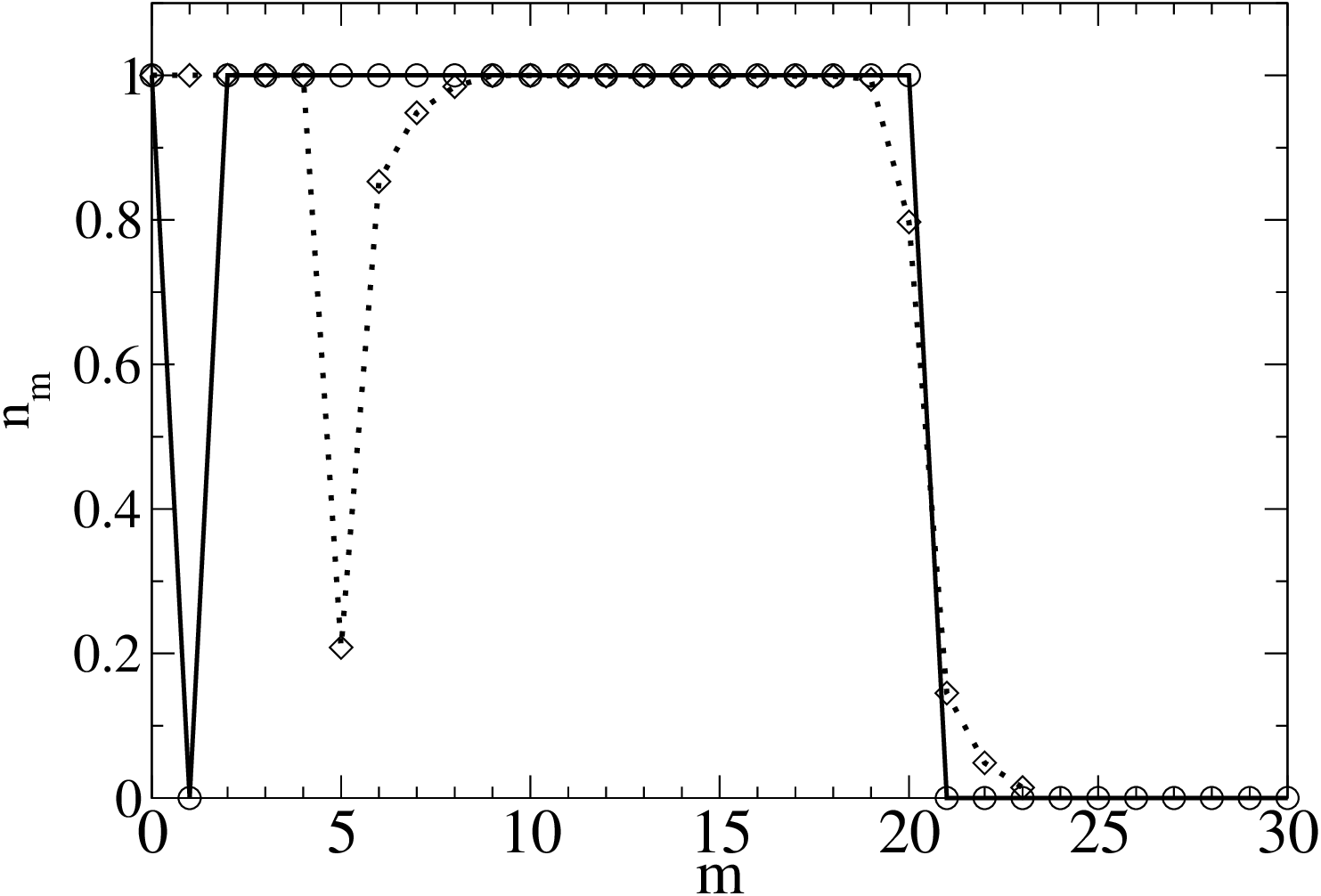}
\caption{\label{fig:one}
Results for the occupation numbers of single particle states $n_m$ in the absence of electron-hole interaction.
Circles are HF results and diamonds are exact result (Solid and dotted lines are guides to the eye).   
Here $\hbar\Omega=3meV$ and $N=20$.
\vspace{0.5in}
}
\end{figure}

Before we present the numerical results  we test  the accuracy of our HF approach.
In the limit where  the electron-hole interaction is absent
electrons and holes are independent of each other and 
exact diagonalization results are  known \cite{Yang3}.
In Fig. \ref{fig:one} we have compared HF occupation numbers of single particle states
with those of  exact results
(In the HF approach occupation numbers are just $n_m=v_m^2$).
The following parameters are  used:  $N=20$,  $\hbar\Omega=3meV$, $\mu^{+}=5meV$, and $B=5.9T$.
We see that when  the effect of quantum fluctuations are included the  
suppression in the occupation numbers become 
more broader and less deeper as a function of $m$.  
Also the positions of  the minimum of the density 
depletion are somewhat different between the two.  Nonetheless we see that HF correctly captures qualitatively
the right physics of reconstruction near the MDD.  The total amount of electron or hole density depletion
in the interior of the droplet  is exactly one in both cases: $\sum_{m=0}^{15}u_m^2=1$.

\subsection{Order parameter  of reconstructed droplet}

\begin{figure}
\includegraphics[width= 0.45  \textwidth]{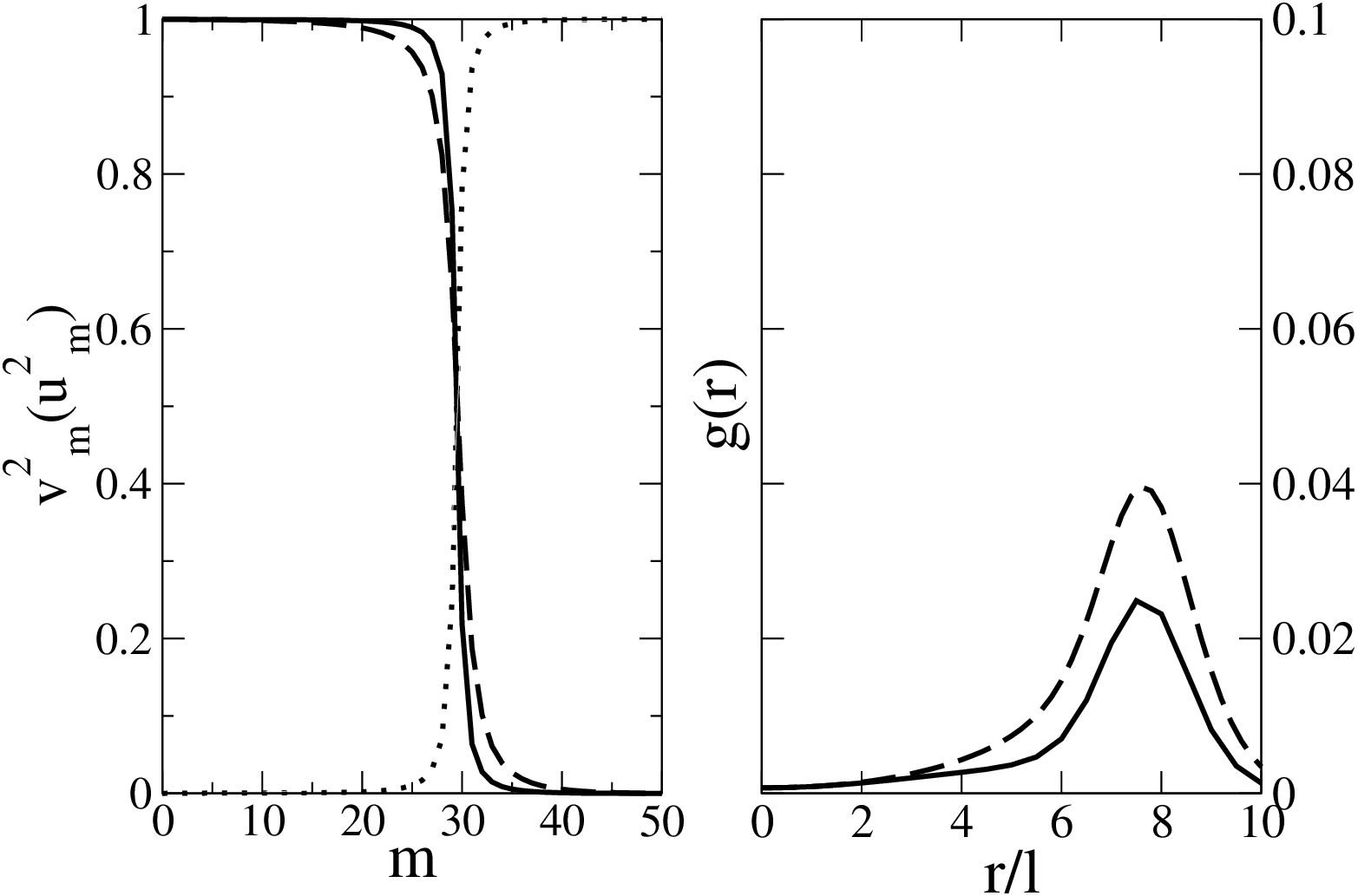}
\caption{ \label{fig:or_1}
Left panel: Particle occupation numbers $v_m^2$ (solid line) 
are plotted as a function of quantum number $m$,  which label single particles states.
$u_m^2=1-v_m^2$ are also plotted  (dotted line). 
Right panel: The order parameter $g(r)$ is plotted as a function of $r$.
Here $\hbar\Omega=2.35$meV and $N=30$. For comparison  $v_m^2$ and $g(r)$  for $d=0$ are also 
shown (dashed lines).}
\vspace{0.5in}
\end{figure}

The order parameter, i. e., the condensate density, is
\begin{eqnarray}
g(r)&=&\langle\Psi_{\uparrow}(r)\Psi_{\downarrow}(r)\rangle\nonumber\\
&=&
\sum_{mm'}\langle c_{m\uparrow}c_{-m'\downarrow}\rangle \phi_m(r)\phi_{-m'}(r),
\end{eqnarray}
where $\Psi_{\uparrow}(r)=\sum c_{m\uparrow}\phi_m(r)$
and $\Psi_{\downarrow}(r)=\sum c_{-m\downarrow}\phi_{-m}(r)$
are electron and hole field operators, respectively.
Since the optical selection rule of excitons \cite{yang} requires $m=m'$ we have
\begin{eqnarray}
g(r)&=&
\sum_{m}\langle c_{m\uparrow}c_{-m\downarrow}\rangle |\phi_m(r)|^2.
\end{eqnarray}
Since  
\begin{eqnarray}
<c_{m\uparrow}c_{-m\downarrow}>=v_mu_m=\frac{1}{2}\frac{\Delta_m}{E_m}.
\end{eqnarray}
we have 
\begin{eqnarray}
\label{order}
g(r)=\sum_{m}  v_mu_m|\phi_m(r)|^2.  
\end{eqnarray}
Note that $|\phi_m(r)|^2$ is peaked at $r\sim \sqrt{2(m+1)}\ell$ with a width $\ell$ (see Eq.\ref{wavef}).

Before we show the results for the order parameter let us first investigate
when the MDD is realized.
Fig. \ref{fig:or_1} displays the calculated $v_m^2$ and $u_m^2$
for
$\hbar\Omega=2.35meV$, 
$N=30$, $d=0.5\ell$, and $B=5T$.
The chemical potential is $\mu^{+}=5.82meV$.
The electron and hole occupation numbers, $v_m^2$, are one except near the edge region $m\approx m_c=29$.
Note that the {\it radius of the edge} of the droplet is $r_c=\sqrt{2(m_c+1)}\ell$.
The electron or hole density of this MDD is  given by
\begin{eqnarray}
n_{e,h}(r)=\sum v_m^2 |\phi_m(r)|^2.  
\end {eqnarray}
This  density  looks approximately like a step function 
\begin{equation}
n_{e,h}(r)\approx
\left\{
\begin{array}{cl}
\frac{1}{2\pi\ell^2},& r< \sqrt{2N}\ell \\
0 , & r> \sqrt{2N}\ell.
\end{array}
\right.
\end{equation}
Thus, in a parabolic trap with equal number of electrons and holes  a {\it uniform} density state 
can be  realized in the strong magnetic field limit with the particle density   $1/2\pi\ell^2$.

Let us  investigate the condensate order parameter for this state.
Near the edge, where $m_c\sim 29$, 
both $v_m^2$ and $u_m^2$ are non-zero, and, consequently, the order parameter $v_mu_m$ is
non-zero, see Fig. \ref{fig:or_1}.  
A rough shape of $g(r)$ may be deduced as follows.  The product $v_mu_m$ is roughly
given by a delta function $\delta_{m,m_c}$.  It then follows from Eq.~(\ref{order}) that
$g(r)\sim |\phi_{m_c}(r)|^2$, which is peaked near the edge $r_c$.
In this MDD  electrons and holes  are surrounded by condensed excitons near the edge of the droplet\cite{paquet}.
We have also calculated $v_m^2$ and  $g(r)$ 
for an increased value for  the strength of electron-hole interaction, corresponding to
$d=0$ (see dashed lines). 
We notice that 
the width of the ring where   the order parameter is non-zero  becomes larger compared to the case of 
$d=0.5\ell$, see the dashed curve in the right panel of Fig. \ref{fig:or_1}.

\begin{figure}
\includegraphics[width=0.45\textwidth]{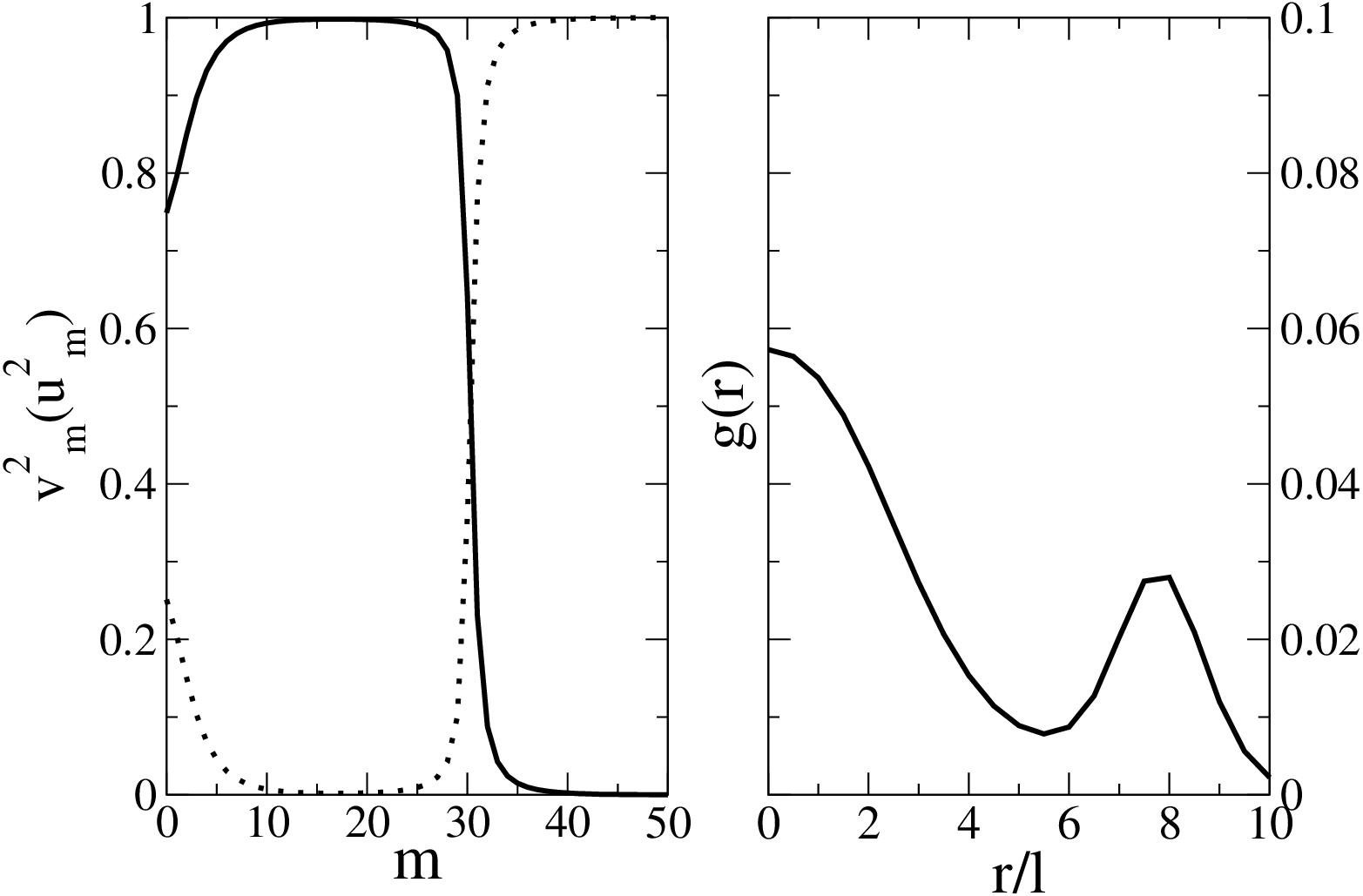}
\caption{ 
\label{fig:or_2}
 Same as in  Fig. \ref{fig:or_1}  but with a smaller value of the strength of the confinement potential
 $\hbar\Omega=2.27meV$.
}
\vspace{0.5in}
\end{figure}

\begin{figure}
\includegraphics[width=0.45\textwidth]{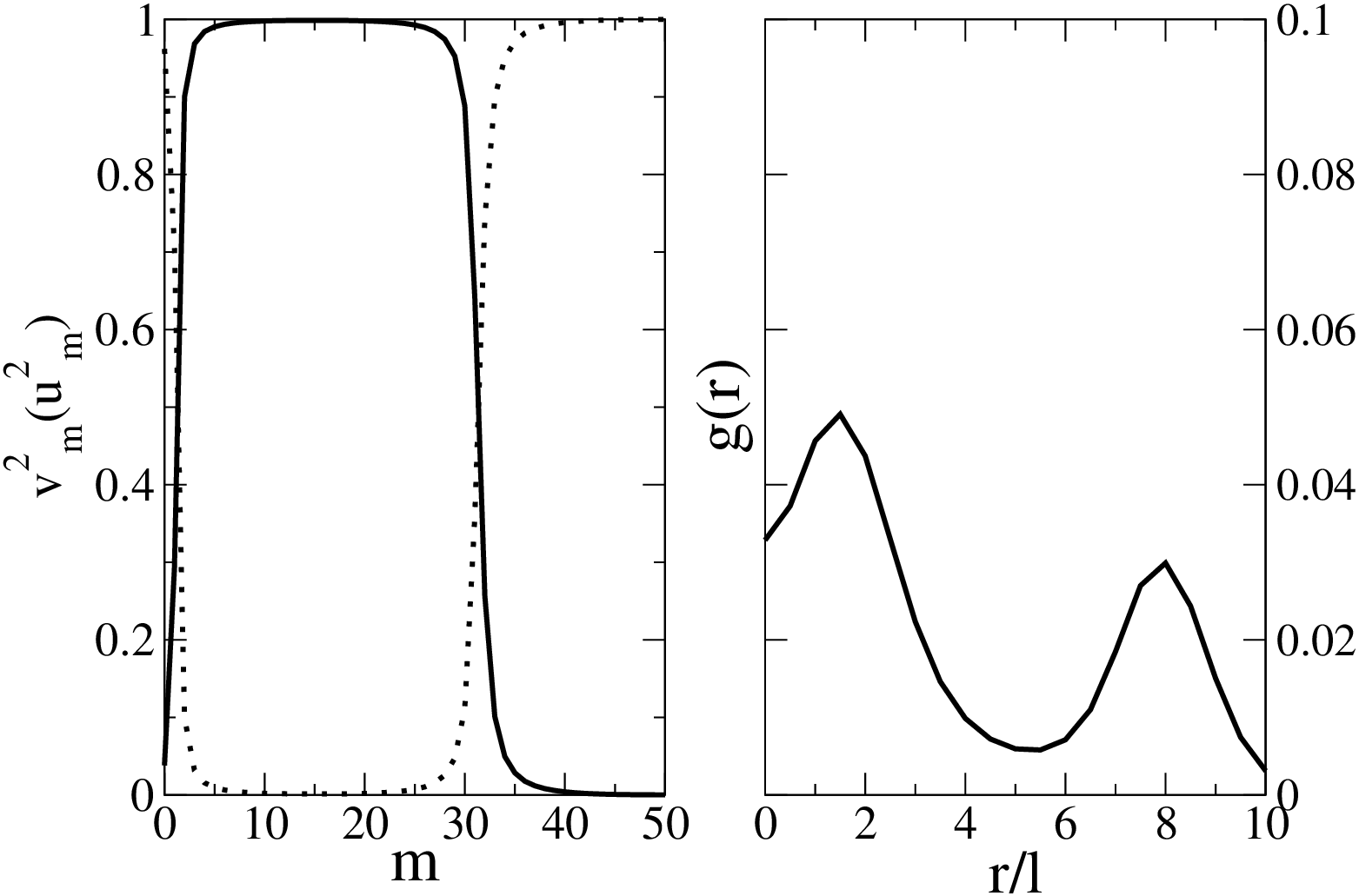}
\caption{\label{fig:or_3} Same as in  Fig. \ref{fig:or_1} but with $\hbar\Omega=2.2meV$.
}
\vspace{0.5in}
\end{figure}

\begin{figure}
\includegraphics[width=0.45\textwidth]{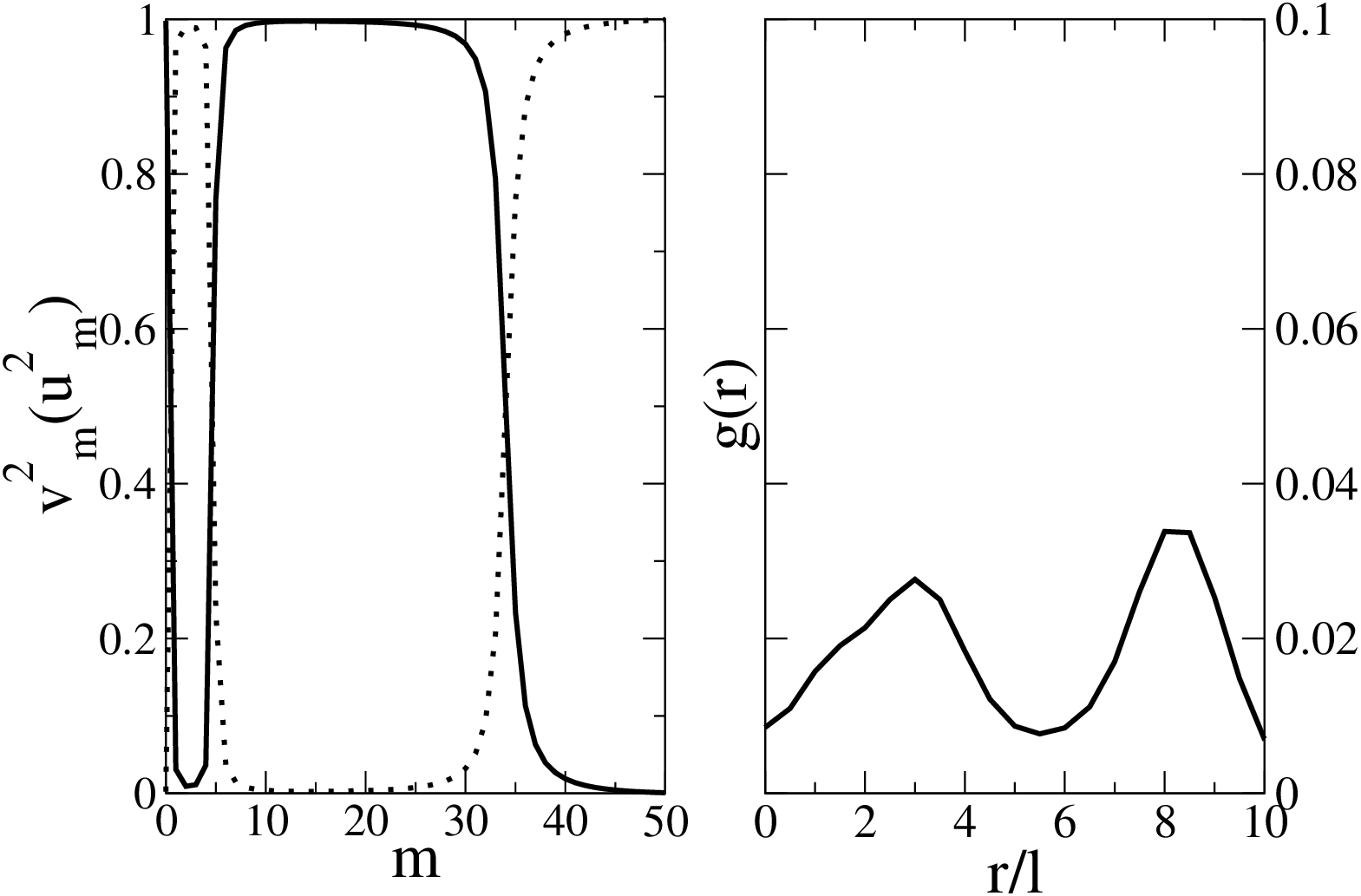}
\caption{\label{fig:or_4}Same as in  Fig. \ref{fig:or_1}  but with $\hbar\Omega=2.0meV$.
}
\vspace{0.5in}
\end{figure}

\begin{figure}
\includegraphics[width=0.45\textwidth]{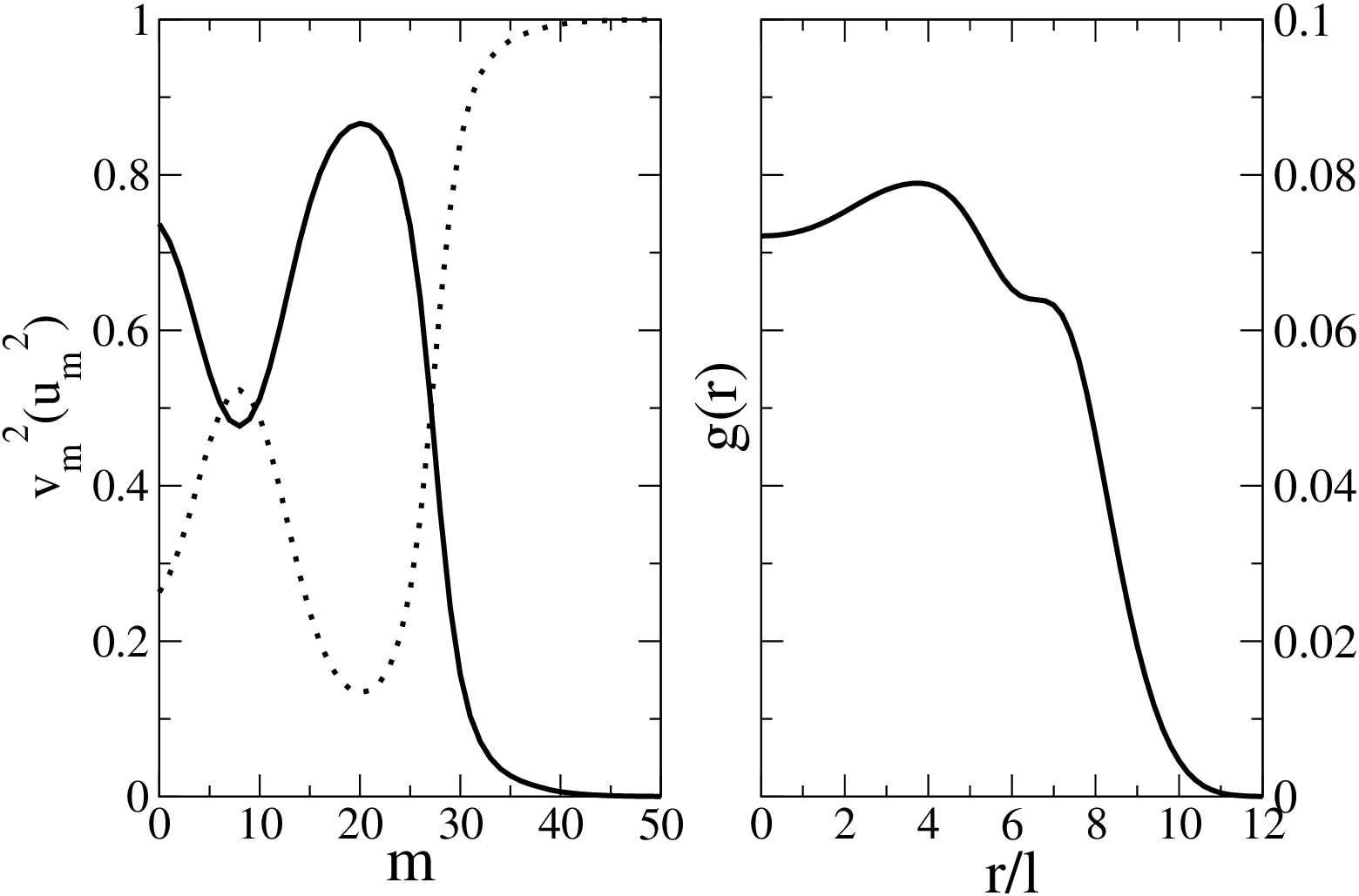}
\caption{\label{fig:or_5}
$v_m^2$ and $g(r)$ for a  non-uniform state are shown.
The parameters are $\hbar\Omega=2meV$ and $N=20$.
}
\vspace{0.5in}
\end{figure}

We now show the results on how the uniform MDD
reconstructs as the strength of the confinement potential {\it decreases}.  We will call these
reconstructed states {\it nearly uniform droplets}.  When a
smaller value  $\hbar\Omega=2.27meV$ is used $v_m^2$ near $m=0$ deviates from one and
$u_m^2$ starts to take non zero values, see  Fig. \ref{fig:or_2} .  The self-consistent value of
the chemical potential is $\mu^+=5.76$meV.  In this case the  order parameter develops
two peaks.
This feature can be understood by investigating the  product of $v_mu_m$: it is non zero near $m=0$ and $m=29$, and
two peaks are expected.
The quantity $\sum_m^{15}u_m^2=0.92$ gives the number of electrons or holes  added  
from the interior of the droplet to the edge.
As  $\hbar\Omega$ is reduced further to $2.2meV$ the value of  $v_m^2$
is almost zero at $m=0$ and   $\sum_m^{15}u_m^2=1.85$ (see  Fig. \ref{fig:or_3} ). 
This means that more particles are added from the interior of the droplet to the edge.
The value of the chemical potential is $\mu^+=5.63$meV.
When $\hbar\Omega$ is reduced
even further to $2.0meV$ the occupation number $v_m^2$ is almost zero, not
at $m=0$, but, at finite
value of $m$ (see  Fig. \ref{fig:or_4} ). In this case the value of the chemical potential is $\mu^+=5.35$meV.
Note that near $m=3$ the occupation numbers $v_m^2$ is nearly zero  while $u_m^2$
takes a maximum value.  The sum is $\sum_m^{15}u_m^2=4.23$. 
We conclude from  results in Figs.  \ref{fig:or_1} - \ref{fig:or_4}  that with decreasing value of $\hbar\Omega$
the total density added  from the interior of the droplet to the edge increases continuously.
Similar trend also exists in the case of electron single dots, but there it changes discontinuously
and takes integer values. 
A  {\it non-uniform} state is also investigated for $\hbar\Omega=2meV$, $\mu^+=3.85$meV, and $N=20$.
Fig. \ref{fig:or_5} displays the calculated $v_m^2$, $u_m^2$ and the order
parameter for this case. The occupation numbers  $v_m^2$ are non-uniform with  the
average value  $1/2$ (For larger values of  $\hbar\Omega$ studied above the average occupation number
was one).
The order parameter is maximum between the center and the edge of the droplet, but does not
exhibit a double peak as before.
We believe that the results of the HF approach are less reliable in describing non uniform states
than nearly uniform states.  The results of Fig.\ref{fig:or_5} should thus  be taken as  rather 
approximate results.

\subsection{Mean field single particle energies}

\begin{figure}
\includegraphics[width=0.45\textwidth]{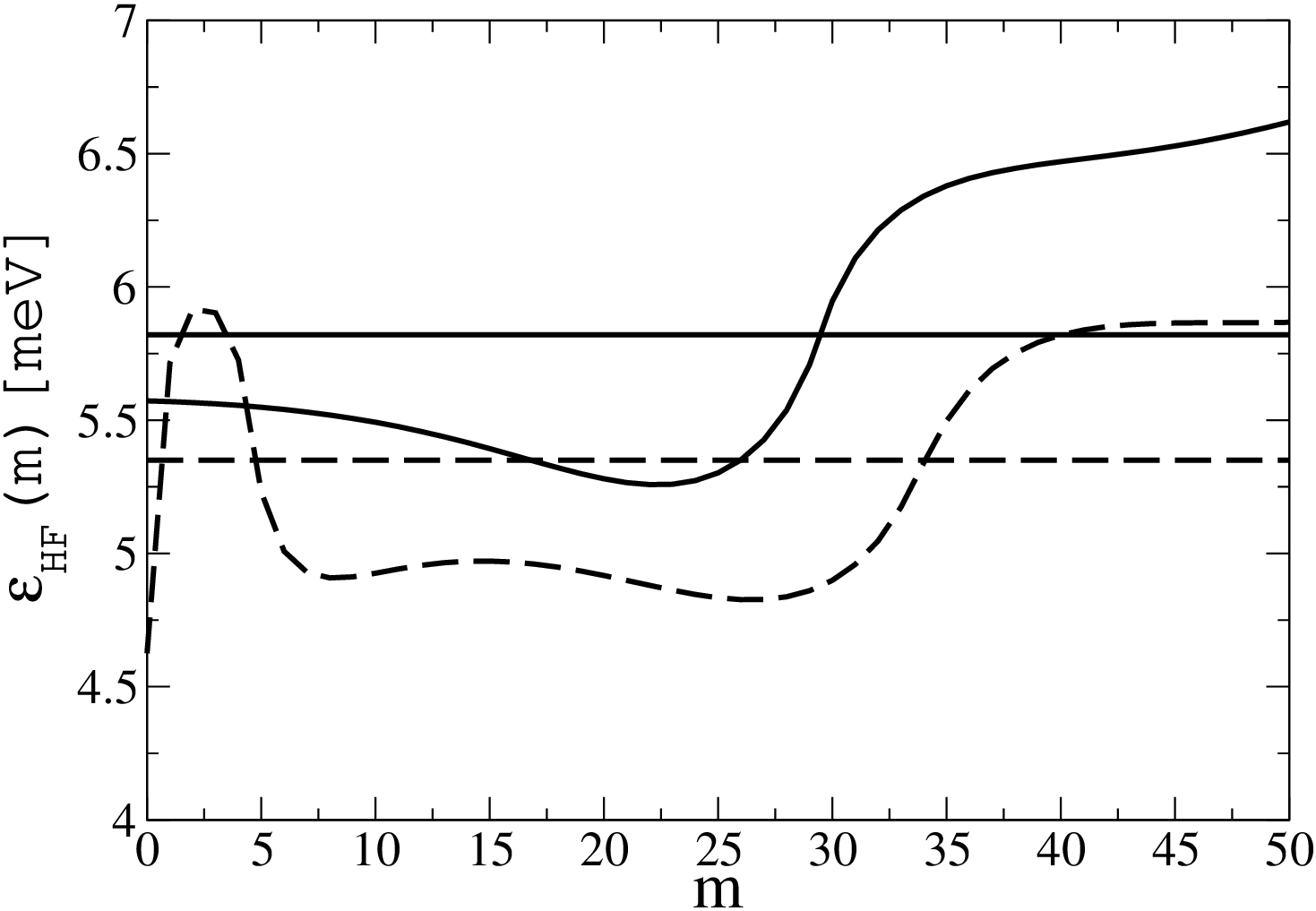}
\caption{\label{fig:self_1}
The effective mean field single particle energies as functions of $m$.
Solid line is for $\hbar\Omega=2.35meV$ ( see Fig. \ref{fig:one}) and the dashed line is
for $\hbar\Omega=2.0meV$ ( see Fig. \ref{fig:or_4}).  For both cases $N=30$.  Horizontal lines represent
the values of the chemical potentials.}  
\vspace{0.5in}
\end{figure}

\begin{figure}
\includegraphics[width=0.45\textwidth]{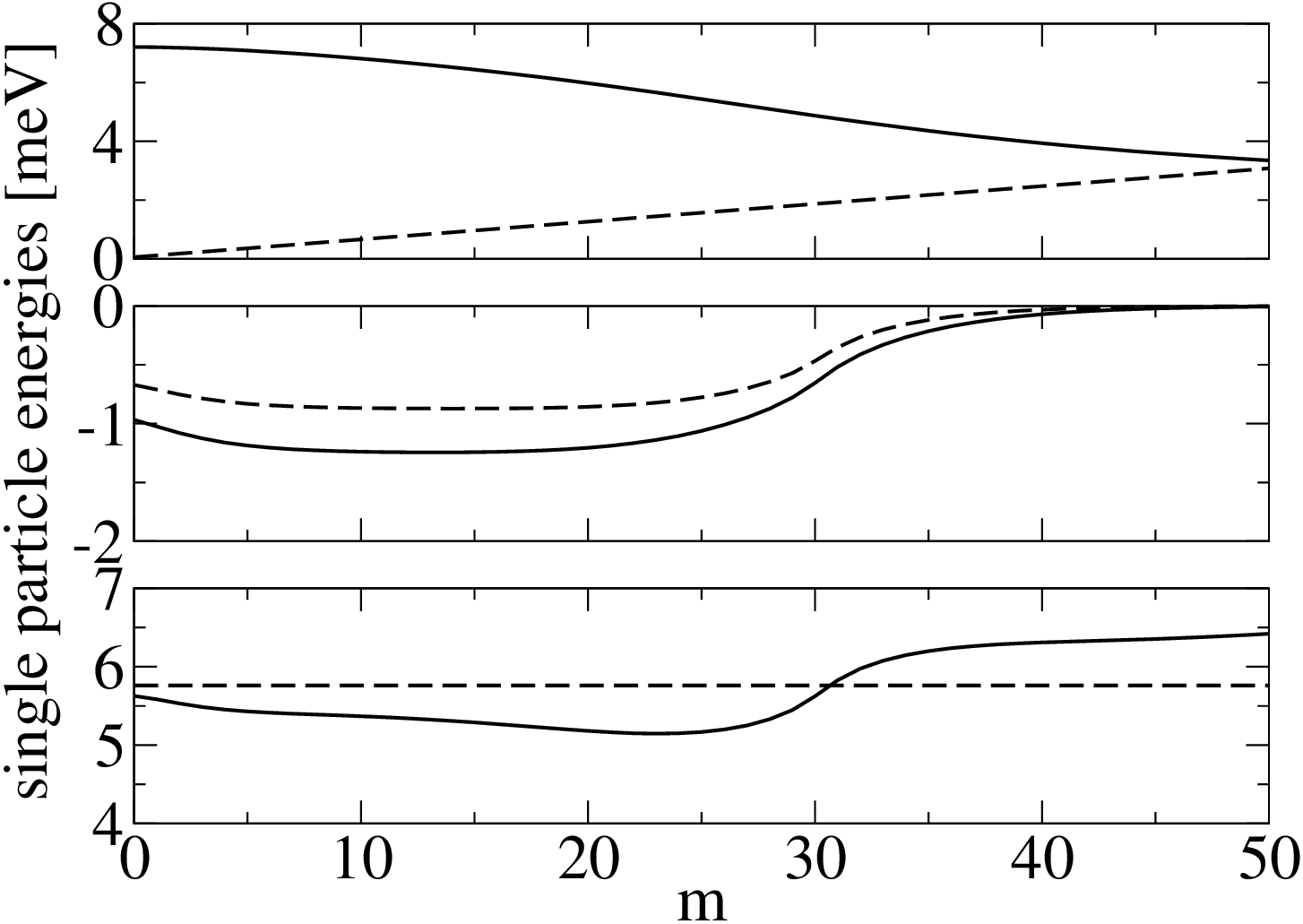}
\caption{\label{fig:self_2} 
Various terms of the mean field single particle energy for $\hbar\Omega=2.27$meV and $N=30$ are plotted
(The corresponding order parameter is plotted in Fig. \ref{fig:or_2}) .
Top panel: Solid line represents the intra layer Hartree self energy, $\Sigma_{H,s}(m)$,  while
dashed line  represents the confinement single particle energy.
Middle panel: Solid line represents the intra layer exchange self energy, $\Sigma_X(m)$, while
dashed line  represents the inter layer Hartree self energy $\Sigma_{H,d}(m)$.
Bottom panel: Solid line represents
the effective mean field single particle energy, $\epsilon^{\sigma}_{HF}(m)$, as functions of $m$
while dashed line represents the chemical potential.
}
\vspace{0.5in}
\end{figure}

\begin{figure}
\includegraphics[width=0.45\textwidth]{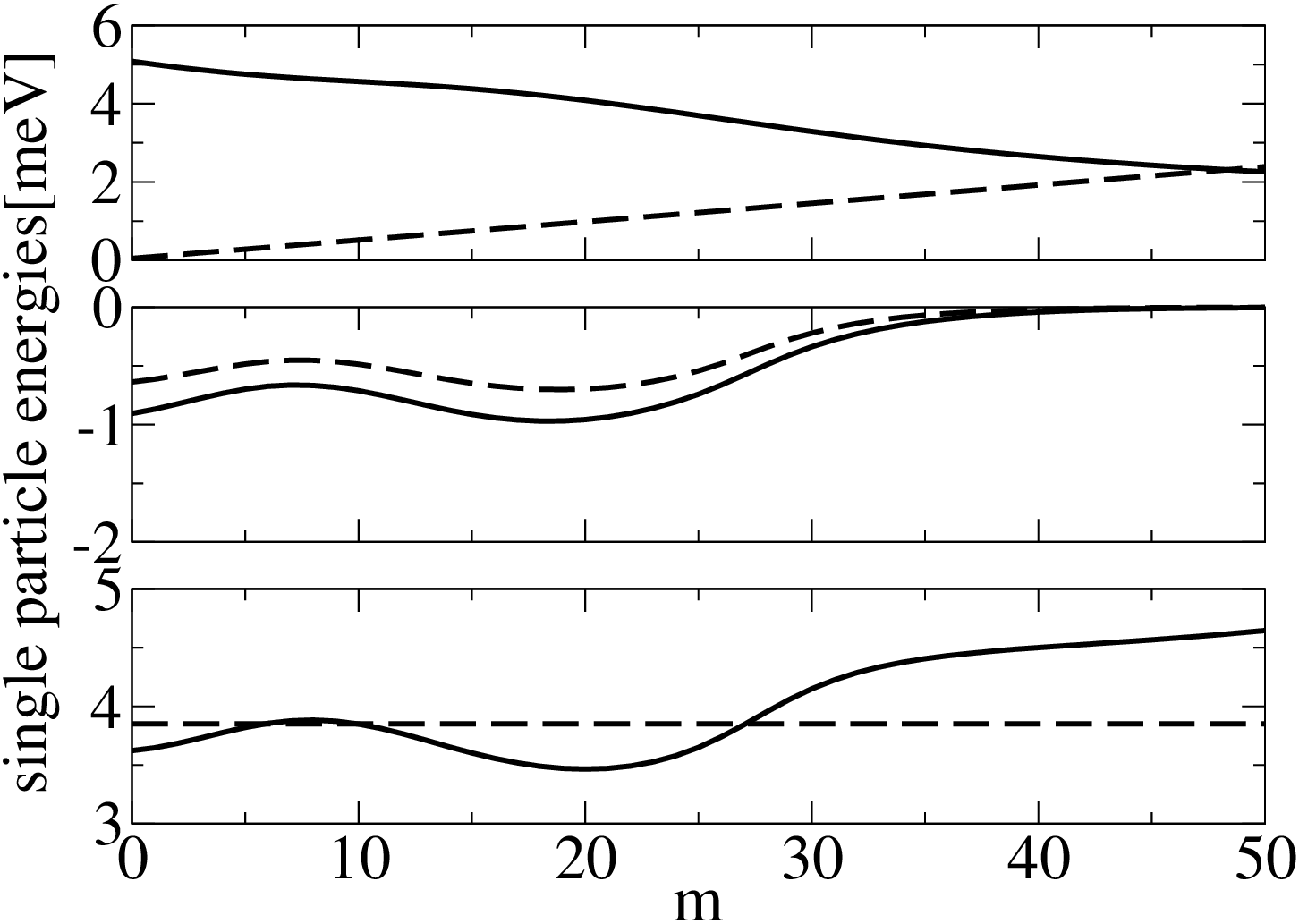}
\caption{\label{fig:self_3}
Same as in  Fig. \ref{fig:self_2} but for the non-uniform state.  
The corresponding order parameter is shown   in  Fig. \ref{fig:or_5}.
}
\vspace{0.5in}
\end{figure}

A strong confinement potential pushes particles in each layer to the center of the potential and 
the particle density of the droplet 
takes  the largest possible value   
except near the edge of such a MDD. 
When the confinement potential becomes weaker some particles move
from inside of the droplet to the edge, i.e., a reconstruction takes place,
and, consequently, the size of the droplet expands in the plane of 2D system.
It is not simple to predict exactly where this density depletion takes place in the droplet since 
it is a consequence of a non-trivial  interplay between the confinement potential, Hartree potential,
exchange potential, and  electron-hole interactions.

The mean field single particle energy $\epsilon^{\sigma}_{HF}(m)$, given
in Eq.~ (\ref{HFenergy}), reflects this interplay to some degree.
A rough shape of $v_m^2$ may be obtained from   the shape of the mean field single particle energy
$\epsilon^{\sigma}_{HF}(m)$:  
In the absence of electron-hole pairing correlation 
single particle states with $\epsilon^{\sigma}_{HF}(m)$ smaller  than
the  chemical potential are  occupied with probability one.  However, since we do have pairing in the BEC 
this is only approximately correct.
The  mean field single particle energies $\epsilon^{\sigma}_{HF}(m)$ are 
plotted for $\hbar\Omega=2.35$meV and $2.0$meV in Fig. \ref{fig:self_1}.
The corresponding $v_m^2$ are shown in  Fig.\ref{fig:or_1} and Fig.\ref{fig:or_4}, respectively.
As the strength of the confinement potential decreases the local maximum of $\epsilon^{\sigma}_{HF}(m)$
moves from the center of the droplet and the width of this local maximum becomes broader. 
The resulting  occupation numbers $v_m^2$ are   approximately consistent with the shape of $\epsilon^{\sigma}_{HF}(m)$.
Even for  a  non-uniform droplet 
the position of the peak    in $\epsilon^{\sigma}_{HF}(m)$  coincides 
with the position of the suppression of $v_m^2$: see the peak near $m=7$ in $\epsilon^{\sigma}_{HF}(m)$
(Fig.\ref{fig:self_3}) and the dip in $v_m^2$ near $m=7$ (Fig. \ref{fig:or_5}).

The various self energies appearing in  $\epsilon^{\sigma}_{HF}(m)$  are plotted 
for  $\hbar\Omega=2.27$meV in Fig. \ref{fig:self_2}.  
The presence of a peak in $\epsilon^{\sigma}_{HF}(m)$ near $m=0$  is related to the peaks in the 
self energies  $\Sigma_X(m)$ and $\Sigma_{H,d}(m)$.
The corresponding  occupation number  $v_m^2$ decreases near $m=0$ , see Fig. \ref{fig:or_2}.
(We notice that $\Sigma_{H,d}(m)$ is almost flat in the region $6<m<20$.
$\Sigma_{H,d}(m)$ displays  qualitatively the
same dependence as the exchange self energy $\Sigma_X(m)$.  
When $d=0$ they are actually identical and are flat for $m<m_c$ \cite{Yang2}).

\section{Conclusions and discussion}

We have investigated  the  shape of  the condensed  magnetoexcitons 
in circular traps in strong magnetic fields.   
Our model consists of a  parabolic confinement in lateral directions and a delta function like
confinement along the perpendicular axis.
We have applied a mean field theory, which  is expected to be qualitatively correct.
We find that as the strength of the confinement potential weakens, or equivalently as $B$ increases,
the uniform state becomes unstable and
density depletion starts to occur in the interior  of the droplet.  
We found that the amount of density depletion increases
continuously with the decrease in the strength of the confinement potential.  
As a consequence of these reconstructions the order parameter changes from displaying
one peak at the edge to displaying  one inner peak  and another peak at the edge
for decreasing confinement strength.  
When density depletions are more severe, ie, when the confinement potential is rather weak, the order parameter
may display one  broad peak.

The reconstruction may be observed experimentally since 
the spatial shape of the order parameter   changes.
The structure in  the order parameter $g(r)$ may be investigated experimentally by observing
angular distribution of photoluminescence since it reflects 
the Fourier transformed order parameter in the momentum space.
An expression for how the PL angular profile depends on  the order parameter
is given explicitly in Keeling et al \cite{Kee}.  They point out that it may provide a diagnostic for the existence
of BEC.

Note that decreasing $\hbar\Omega$ and increasing $B$ has the same effect.
Experimentally  either of these two parameters may be varied.
This can be understood as follows.  The strength of the confinement strength
enters as a dimensionless parameter in units of the Coulomb energy scale:
$\tilde{\gamma}_{\sigma}=\gamma_{\sigma}/(e^2/\epsilon\ell)$.
The reconstruction of the shape of the droplet will depend  on this parameter, which is
a function of $B$ and  $\hbar\Omega$:  $\tilde{\gamma}_{\sigma}\sim (\hbar\Omega)^2/B^{3/2}$.  
It may be easier experimentally to change $B$  than $\hbar\Omega$.

Also it is desirable to increase the particle numbers one order of magnitude and 
investigate the shape of the 
condensate droplet.   However, this is a non-trivial task since the matrix elements of the 
particle-particle interactions are difficult to calculate numerically when the quantum number $m$ are large.

We comment on  effects that are not investigated in this paper. 
It would be interesting to study  how 
the corrections to the Hartree Fock theory affect quantitative aspects of  
the shape of the order parameter.
This 
can be investigated by performing numerical exact diagonalization.  
Finite quantum well width corrections of the electron and hole wavefunctions
will weaken intralayer interactions and hence favor excitonic states over  FQHE
states.  Realistic valence band  structure effects \cite{yang}  can be included through the matrix elements of the
intra and inter layer Coulomb interaction.  Corrections to the lowest Landau level approximation  may have 
quantitative relevance \cite{Si}.

This work is supported by Korea Research Foundation   
grant KRF-2003-015-C00223, and  by grant No.(R01-1999-00018) from the interdisciplinary
research program of the KOSEF.


\end{document}